\documentstyle[12pt,epsf,rotate,psfig]{article}
\pssilent
\makeatletter
\title{Evidence of a Change in the Long Term Spin-down Rate of the X-ray Pulsar 4U 1907+09}
\author{A. Baykal$^1$, S.\c{C}. \.{I}nam $^2$, E. Beklen $^{1,3}$ \\ $^1$ Physics Department,\\ Middle East Technical University, 06531 Ankara, Turkey \\ altan@astroa.physics.metu.edu.tr \\ $^2$ Department of Electrical and Electronics Engineering,\\ Ba\c{s}kent University, 06530 Ankara, Turkey \\ inam@baskent.edu.tr \\ $^3$ Physics Department, \\ S\"{u}leyman Demirel University, 32260 Isparta, Turkey \\ elif@astroa.physics.metu.edu.tr}
\date{}
\begin{document}
\maketitle
\begin{abstract}
We analyzed RXTE archival observations of 4U 1907+09 between 17 February 1996 and 6 March 2002. The pulse timing analysis showed that the source stayed at almost {\bf{constant}} period around August 1998 and then started to spin-down at a rate of $(-1.887\mp 0.042)\times 10^{-14}$ Hz s$^-1$  which is $\sim$ 0.60 times lower than the long term ($\sim 15$ years) spin-down rate (Baykal et al. 2001). Our pulse frequency measurements for the first time resolved significant spin-down rate variations since the discovery of the source. We also presented orbital phase resolved X-ray spectra during two stable spin down episodes during November 1996 - December 1997 and March 2001 - March 2002. The source has been known to have two orbitally locked flares. We found that X-ray flux and spectral parameters except Hydrogen column density agreed with each other during the flares.We interpreted the similar values of X-ray fluxes as an indication of the fact that the source accretes not only via transient retrograde accretion disc (in't Zand et al. 1998) but also via the stellar wind of the companion (Roberts et al. 2001), so that the variation of the accretion rate from the disc does not cause significant variation in the observed X-ray flux. Lack of significant change in spectral parameters except Hydrogen column density was interpreted as a sign of the fact that the change in the spin-down rate of the source was not accompanied by a significant variation in the accretion geometry.

\end{abstract}
   	      
\section{Introduction}
The X-ray source 4U 1907$+$09 is an accretion powered X-ray pulsar accreting plasma from a blue supergiant companion star. Since its discovery in the Uhuru survey (Giacconi et al. 1971), it has been observed by various X-ray observatories including  Ariel V (Marshall \&
Ricketts 1980), Tenma (Makishima et al. 1984), EXOSAT (Cook \&
Page 1987), Ginga (Makishima \& Mihara 1992; Mihara 1995), BeppoSAX (Cusumano et al. 1998), XMPC (Chitnis et al. 1993), IXAE 
(Mukerjee at al.2001) and RXTE (in't Zand, Baykal \& Strohmayer 1998a; in't Zand, Strohmayer \& Baykal 1997, 1998b; Roberts et al. 2001 ;Baykal et al. 2001). Marshall \& Ricketts (1980) determined the orbital period to be 8.38d using Ariel V observations. They also found two flares, namely a primary and a secondary {\bf{occuring near periastron and apastron}}, and EXOSAT and
RXTE observations (in't Zand et al. 1998a,b) have clearly shown
that these flares are locked to orbital phases separated by half an
orbital period. Using Tenma
observations, Makishima et al. (1984) also found the pulse period of the source to be 437.5 sec. Makishima et al. (1984) and Cook \& Page (1987)
discussed the possibility that the two flares are caused by an equatorial disc-like envelope (circumstellar disc) around the companion star. When the neutron star crosses the disc, the mass accretion rate onto the neutron star increases temporarily. We expect to see two peaks while the neutron star passes twice through the circumstellar disc of the companion star in every cycle. We also expect a transient accretion disc formation around the neutron star while the neutron star passes through this circumstellar disc. For the formation of the circumstellar disc around the companion star, the companion must most likely be a Be type star. Recent optical observations showed that  the optical companion of 4U 1907$+$09 is an O8/O9 supergiant with a dense stellar wind (Cox, Kaper, Mokiem 2005). Formation of a transient accretion disc around the neutron star as a result of disruption of the dense stellar wind of a massive companion is quite possible as well (Blondin et al. 1990). 

The timing measurements from EXOSAT (Cook \& Page 1987) and
RXTE observations (in't Zand et al. 1998a,b; Baykal et al. 2001) showed continuous spin down of the neutron star.The source was found to be spinning down almost at a rate of ${\dot \nu}=(-3.54 \pm 
0.02)\times {10^{-14}} {Hz s^{-1}}$ for more than 15 years (Baykal et al. 2001).     
Timing analysis of RXTE (in't Zand et al. 1998) and IXAE (Mukerjee et al. 2001) observations of 4U 1907-09 also revealed transient oscillations with periods  of about 18.2s and 14.4s respectively. These transient oscillations {\bf{were}} 
suggested to be related to the presence of a transient retrograde accretion disc which is also responsible for the slowing down of the pulsar. 

In this paper we present the results of timing and spectral studies of archival RXTE-PCA observations of 4U 1907+09,  and report a change in the spin down rate of this system.

\section{Instrument and Observations}
The Proportional Counter Array (PCA)(Jahoda et al.1996) on board the RXTE consists of five identical proportional counter units (PCUs) coaligned to the same point in the sky, with a total effective area of approximately 6250 ${cm}^2$, and a field of view of $\sim 1 ^{0}$ FWHM, operating in the 2-60 keV energy range. The number of active PCUs is varied between one and five during the observations.Observations after 2000 May 13 belong to the observational epoch for which background level for one of the PCUs (PCU0) increased due to the fact that this PCU started to operate without a propane layer. Latest combined background models (CMs) were used together with FTOOLS 6.0 to estimate the appropriate background.    

The RXTE-PCA observations used in this work are listed in Table 1.

\begin{table}[hb]
\caption{Observation list for 4U 1907+09}
\begin{center}
\begin{tabular}{c c c}\hline
Time of Observation & XTE Prop. ID & Number of \\
start       end   &     & Observations \\
dd mm yyyy dd mm yyyy     &   &   \\ \hline \hline
17 02 1996-23 02 1996  & 10155 & 8 \\
25 11 1996-14 12 1997 & 10154-20146  & 24 \\
26 07 1998-01 10 1998  & 30093 & 14 \\
10 03 2001-06 03 2002  & 60061 & 38 \\
\hline \hline
\end{tabular} 
\end{center}
\end{table}

\section{Analysis and Results}

\subsection{Timing Analysis}
In order to obtain background subtracted light curves, 
we generated the background using
the background estimator models based on the rate of very large solar 
events, spacecraft activation and cosmic X-ray emission with the standard PCA
analysis tools. Then we subtracted background light curves from the source light curves obtained from the event data. The background subtracted light curves were also corrected to the barycenter of the solar system and to the binary orbital motion of 4U 1907+09 around its companion. For the correction of binary orbital motion, we used the orbital parameters deduced by In 't Zand, Baykal $\&$ Strohmayer (1998a). In the timing analysis we fold the light curve outside the intensity dips from long data string around 7-10 ksec on statistically independent trial
periods (Leahy et al. 1983). A template pulse profile was constructed
by folding the data on the period giving the maximum $\chi^2$.
Template pulse profile was represented by its Fourier harmonics (Deeter \& Boynton 1985) and cross-correlated with the harmonic representation of average pulse profiles from each observation. The pulse phases obtained from the cross correlation analysis are presented in Figure 1.

\begin{figure}[t]
\begin{center}
\centerline{\psfig{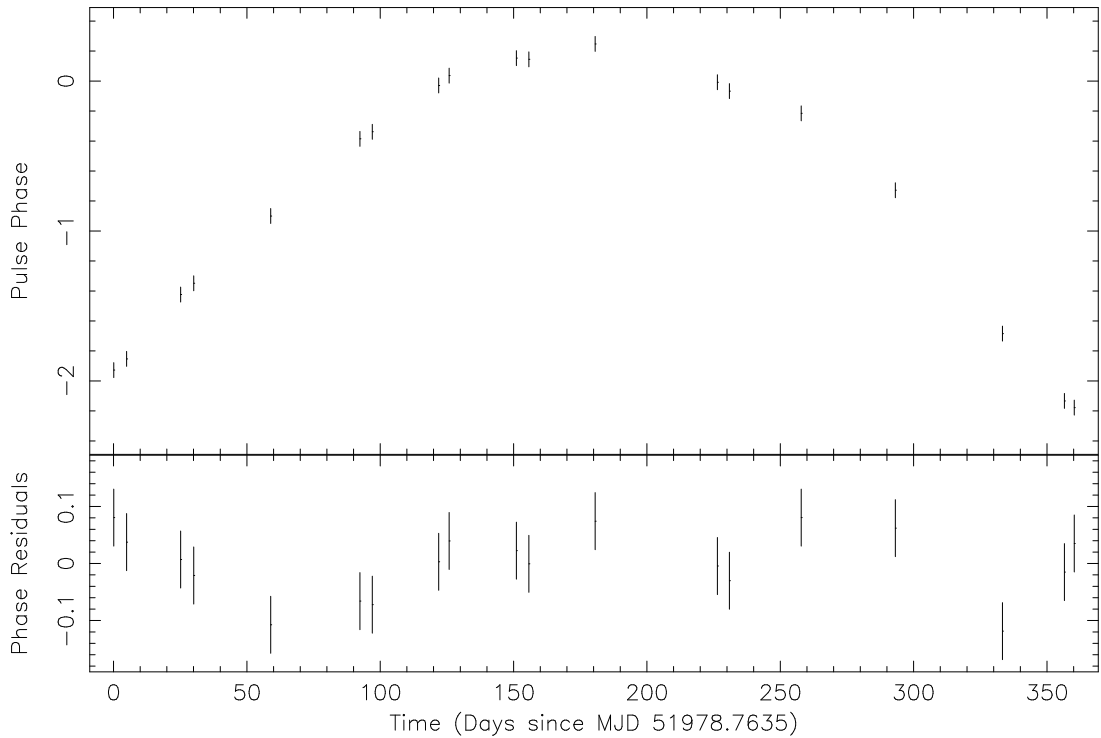}}
\end{center}
\caption{Pulse phase and its residuals fitted to the orbital model presented in Table 2.}
\end{figure}

In order to estimate pulse frequency derivatives, the pulse phases are fitted to the quadratic polynomial
\begin{equation}
\delta \phi = \phi_{o} + \delta \nu (t-t_{o})
+ \frac{1}{2} \dot \nu (t-t_{o})^{2}
\end{equation}
where $\delta \phi $ is the pulse phase offset deduced from the pulse
timing analysis, $t_{o}$ is the mid-time of the observation, $\phi_{o}$ is
the phase offset at t$_{o}$, $\delta \nu$ is the deviation from the mean
pulse frequency (or additive correction to the pulse frequency), and $\dot
\nu $ is the pulse frequency derivative of the source.
The pulse phases and
the residuals of the fit after the removal of the quadratic polynomial
are also presented in Figure 1. Table 2 presents the timing solution of 4U 1907+09.

\begin{table}[t]
\caption{Timing solution of 4U 1907+09 $^a$}
\begin{center}
\begin{tabular}{c|c}\hline \hline
{\bf{$T_{\pi /2}$ (Orbital Epoch) (MJD)}} & $50134.76\mp 0.06$ $^b$  \\ 
$P_{orb}(d)$ & $8.3753\mp 0.0001$ $^b$  \\ 
$a_{x}$sin i (lt s) & $83\mp 2$ $^b$   \\
e & $0.28\mp 0.04$ $^b$ \\
w & $330\mp 7$ $^b$ \\
$t_{0}$ (MJD) $^{c}$ & $52154.6183\mp 0.0005$  \\
Pulse Period (s) & $441.0932\mp 0.0003$ \\
Pulse Period derivative (s s$^-1$) & $(3.653\mp 0.081)\times 10^{-9}$ \\
Pulse freq. derivative (Hz s$^-1$) & $(-1.887\mp 0.042)\times 10^{-14}$ \\ \hline \hline
\end{tabular}
%\multicolumn{2}{l}
\end{center}
$^a$ {\bf{Confidence intervals are quoted at 1$\sigma$ level.}} \\
$^b$ Orbital parameters are taken from In't Zand et al.(1998a).P$_{orb}$=orbital period,{a$_{x}$}sin 
i=projected semimajor axis, e=eccentricity, w=longitude of periastron. \\
$^c$ {\bf{Mid-time of the observation used in pulse timing (see Equation 1 in the text)}} 
\end{table}

The pulse frequency derivative from the sequence of 19 pulse phases
spread over 360 days
was measured as $\dot \nu = (-1.887 \pm  0.042) \times
10^{-14}$ Hz s$^{-1}$. This value approximately $\sim$ 0.60 times lower
then previous measurements of pulse frequency derivatives which was 
measured over 383 days times span {\bf{(Baykal et al. 2001)}}. In this and the 
previous observations Baykal et al., (2001) time span of the observations are 
similar ($\sim $ over year) and in both observations {\bf{possess}} low timing noise. However {\bf{the}}
spin down rate is lowered by  a factor of 0.6 times in latter 
observations.

In order to see the spin down trends clearly
we estimate the pulse frequency histories by taking the derivatives of each pairs of pulse phases from the pulse phases of this work and previous work by Baykal et al., (2001) and presented in Table 3 and Figure 2 along with previous frequency measurements.

\begin{figure}[t]
\begin{center}
\centerline{\psfig{file=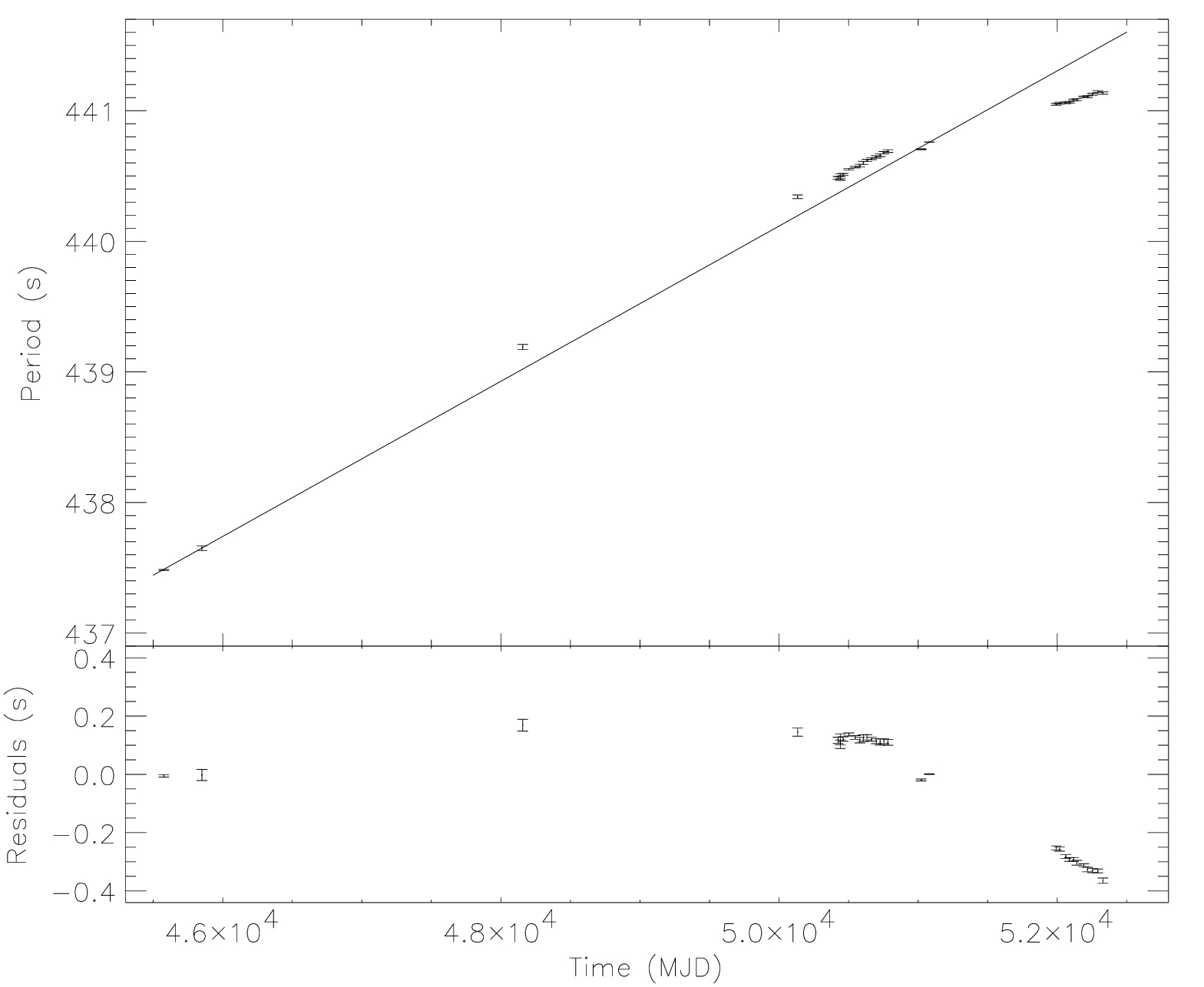,width=\textwidth}}
\end{center}
\caption{Plot of pulse period history listed in Table 3 (Makishima et al. 1984, Cook \& Page 1987, Mihara 1995, In't Zand et al. 1998, Baykal et al. 2001 and this work). Solid line and the residuals correspond to the previous spin-down rate found by Baykal et al. (2001).}
\end{figure}

\begin{table}[hb]
\caption{RXTE pulse period measurements of 4U 1907+09}
\begin{center}
\begin{tabular}{c c c}\hline
Epoch & Pulse Period & Reference \\
(MJD) & (s) &   \\ \hline \hline
45576 & 437.483$\pm$0.004 & Makishima et al.1984 \\
45850 & 437.649$\pm$0.019 & Cook \& Page 1987 \\
48156.6 & 439.19$\pm$0.02 & Mihara 1995  \\
50134 & 440.341$\pm$0.014 & In't Zand et al.1998 \\
50424.3 & 440.4854$\pm$0.0109 & This work\\
50440.4 & 440.4877$\pm$0.0085 & Baykal et al.2001 \\
50460.9 & 440.5116$\pm$0.0075 & This work\\
50502.1 & 440.5518$\pm$0.0053 & This work\\
50547.1 & 440.5681$\pm$0.0064 & This work\\
50581.1 & 440.5794$\pm$0.0097 & This work\\
50606.0 & 440.6003$\pm$0.0115 & This work\\
50631.9 & 440.6189$\pm$0.0089 & This work\\
50665.5 & 440.6323$\pm$0.0069 & This work\\
50699.4 & 440.6460$\pm$0.0087 & This work\\
50726.8 & 440.6595$\pm$0.0105 & This work\\
50754.1 & 440.6785$\pm$0.0088 & This work\\
50782.5 & 440.6910$\pm$0.0097 & This work\\
51021.9 & 440.7045$\pm$0.0032 & Baykal et al.2001 \\
51080.9 & 440.7598$\pm$0.0010 & Baykal et al.2001 \\
51993.8 & 441.0484$\pm$0.0072 & This work\\
52016.8 & 441.0583$\pm$0.0071 & This work\\
52061.5 & 441.0595$\pm$0.0063 & This work\\
52088.0 & 441.0650$\pm$0.0063 & This work\\
52117.4 & 441.0821$\pm$0.0062 & This work\\
52141.2 & 441.0853$\pm$0.0082 & This work\\
52191.4 & 441.1067$\pm$0.0046 & This work\\
52217.2 & 441.1072$\pm$0.0077 & This work\\
52254.3 & 441.1259$\pm$0.0074 & This work\\
52292.0 & 441.1468$\pm$0.0065 & This work\\
52328.8 & 441.1353$\pm$0.0090 & This work\\
\hline \hline
\end{tabular}
\end{center}
\end{table}

The whole dataset of 4U 1907+09 mentioned in previous sections was also analyzed for possible detection of any transient oscillations around $\sim 14-18$s,  similar to those reported from previous RXTE (in't Zand et al. 1998a) and IXAE (Mukerjee at al.2001) observations. In the power spectra of 3-25 keV RXTE-PCA light curve, we searched for transient oscillations around $\sim 0.06$ Hz in the 0.50s binned $\sim 170$ light curve segments of length of 512s. In order to test the significance of these oscillations, we used the approach of van der Klis (1989). As the result of this search, we found only 3 512s segments in the whole dataset with the transient oscillations of significance exceeding 5$\sigma$ at MJD $\sim 50136$, $\sim 51079$ and $\sim 51087$. Our results about transient oscillations are marginal, and therefore inconclusive. Future observations may be useful to observe and analyze possible transient oscillations of the source.    

\subsection{Analysis of the Energy Spectrum}
The complete dataset of 4U 1907+09 mentioned in Section 2.1 were also analyzed to investigate whether there is any significant spectral evolution accompanied with the change in the spin down rate of the pulsar or not. Spectra, background and response matrix files were created using FTOOLS 6.0 data analysis software. Background spectra were generated using the background estimator models based on the rate of very large events, spacecraft activation and cosmic X-ray emission. 

Spectral analysis and error estimation of the spectral parameters were performed using XSPEC version 11.3.2. We extracted individual 3-25 keV PCA spectra each corresponding to a single pulse period. For these spectra, energies lower than 3 keV were ignored due to uncertainties in background modelling while energies higher than 25 keV were ignored as a result of poor counting statistics. In order to take into account the X-ray background from the Galactic ridge and the nearby supernova remnant W49B, we followed the approach used by Roberts et al. (2001): We extracted overall "dip" state spectra of the source for each of the proposal IDs listed in Table 1 and the corresponding background spectra. We found that the count rate and the spectral shape of the dip state spectra were consistent with the models of the diffuse emission from the Galactic ridge (Valinia \& Marshall 1998). So, these dip state observations were used as additional background spectra in spectral fits of the source spectrum. These additional background spectra were created by subtracting the corresponding dip state model background spectrum from the dip state source spectrum using "mathpha" FTOOL. 

\begin{figure}[t]
\begin{center}
\centerline{\psfig{file=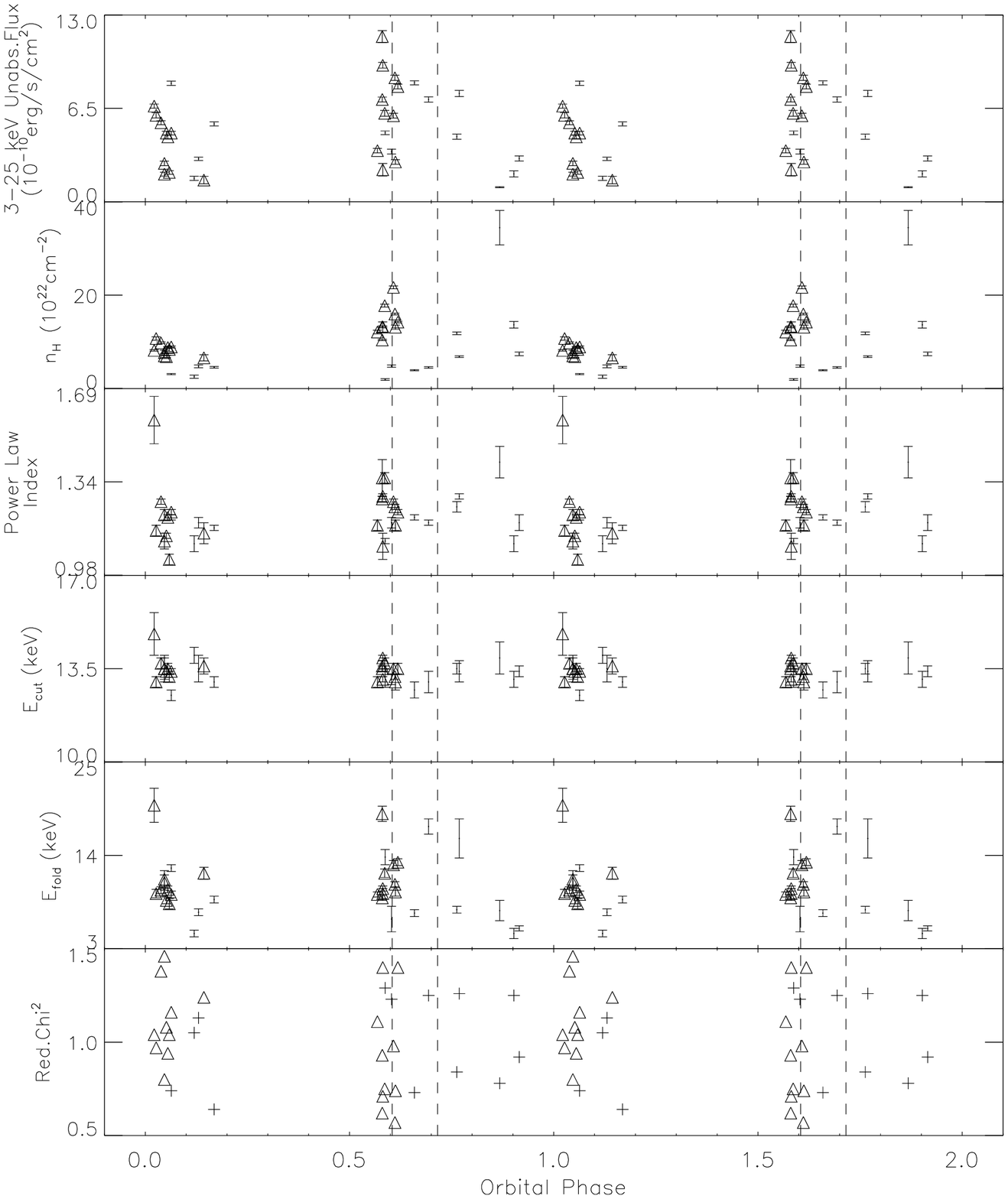,width=14cm,height=19cm}}
\end{center}
\caption{Evolution of 3-25 keV unabsorbed flux, Hydrogen column density, power law index, cut-off energy, e-folding energy, and reduced $\chi^2$ in orbital phase. Vertical dashed lines indicate the orbital phase corresponding to periastron passages. Points marked with triangles indicate observations between March 2001 and March 2002 (proposal ID 60061) for which spin-down rate of the source was lower. The dotted lines indicate $\pm\sigma$ from the periastron.}
\end{figure}

A power-law model with low-energy absorption (Morrison \& McCammon 1983), multiplied by an exponential high energy cut-off function (White, Swank \& Holt 1983) was used to model 3-25 keV spectra of the source. We found no evidence of a statistically significant iron line emission at $\sim 6.4$ keV which is consistent with the previous RXTE spectral fits (Roberts et al. 2001). An additional cyclotron resonance absorption line (Makishima et al. 1990) component at $\sim 19$ keV was required to fit the spectra. The parameters of cyclotron absorption line was found to be consistent with the fundamental cyclotron line parameters found from BeppoSAX observations of the source (Casumano et al. 1998). Obtaining the spectral fits, we {\bf{fixed}} the cyclotron absorption line energy at 19 keV.  The result of the spectral fits as a function of orbital phase were presented in Figure 3. Sample model parameters were listed in Table 4.

\begin{table}
\caption{Sample Spectral Parameters of Individual PCA Observations of 4U 1907+09}
\begin{center}
\begin{tabular}{|c|c|c|}\hline \hline
Parameter & MJD 50547 & MJD 52061 \\ \hline
Hydrogen Column Density ($10^{22}$cm$^{-2}$) & $5.51\mp 0.50$ & $9.13\mp 0.23$ \\
Power Law Photon Index & $1.20\mp 0.04$ & $1.21\mp 0.03$ \\
Power Law Norm. ($10^{-2}$.cts.cm$^{-2}$.s$^{-1}$) & $4.04\mp 0.09$ & $2.83\mp 0.15$ \\
Cut-off Energy (keV) & $12.9\mp 0.2$ & $13.7\mp 0.3$ \\
E-folding Energy (keV) & $7.0\mp 0.5$ & $9.6\mp 0.5$ \\
Cyclotron Depth & $0.14\mp 0.05$ & $0.20\mp 0.02$ \\
Cyclotron Energy (keV) & 19.0 (fixed) & 19.0 (fixed) \\
Cyclotron Width (keV) & $1.12\mp 0.10$ & $1.71\mp 0.34$ \\
3-25 keV Unabsorbed X-ray Flux & $6.36\mp 0.14$ & $4.69\mp 0.25$ \\
 ($10^{-10}$ergs.cm$^{-2}$.s$^{-1}$) & & \\  
Reduced $\chi^2$ (43 d.o.f.) & 0.74 & 0.98 \\
\hline \hline
\end{tabular}
\end{center}
\end{table}

\section{Conclusion}

The spin-down rate of 4U 1907+09 obtained from the pulse arrival times between March 2001 and March 2002 was found to be $\sim$ 0.60 times lower than both the previous RXTE measurements of spin-down rate between November 1996 - December 1997 by Baykal et al., (2001) and the long term spin down trend of the source between 1983 and 1997 (See Figure 1 in Baykal et al. 2001).  In this and the previous RXTE observations, time span of the observations were similar ($\sim $ over year) and in both observations posses low timing noise. However spin down rate is lowered by a factor of 0.6  in latter observations. It is also interesting to observe that the spin-down rate is consistent with zero around MJD 51000 before this significant change in spin-down rate. Our pulse frequency measurements for the first time resolved significant spin-down rate variations since the discovery of the source.

The steady spin-down rate (Baykal et al. 2001) and the presence of occasional transient oscillations (in't Zand et al. 1998a; Mukerjee et al. 2001) in 4U 1907+09 support the idea that the source accretes from retrograde transient accretion disc. The possibility of a prograde disc for which the magnetospheric radius should be close to the corotation radius so that the magnetic torque overcomes material torque is not likely to be the case for this system (in't Zand et al. 1998a). If the disc is retrograde and the material torque is dominant, using Ghosh\& Lamb (1979) accretion disc model, a decrease in spin-down rate of the neutron star should be a sign of a decrease in the mass accretion rate coming from the accretion disc. If the disc accretion is the only accretion mechanism, this decrease in mass accretion rate should also lead to a decrease in X-ray flux of the system.

From the X-ray spectral analysis of the source, we found no clear evidence of a correlation between 3-25 keV flux and the change in spin-down rate (see Figure 3). The flux levels were about the same for latter observations with low spin-down rate compared to the rest of the observations. This was not our expectation if we only consider disc accretion and may be because of the fact that the total mass accretion rate is not only due to only disc accretion, but also accretion from stellar wind may contribute to the total mass accretion rate (see also Roberts et al. 2001) . In case of accretion from both transient disc and wind, the change in mass accretion rate from the accretion disc might not cause a substantial change in X-ray flux. However, it is important to note that the orbital coverage of the latter (proposal ID 60061) observations is poor and limited to the phase locked flares in orbital phases $\sim 0.05$ and $\sim 0.6$ as seen in Figure 3. Future observations of the source may be helpful to have a better understanding of a possible relation between spin-down rate and X-ray flux.

Examining spectral parameters plotted in Figure 3 especially for flare parts, we also found an evidence of an increase in Hydrogen column density ($n_H$) for the latter observation with lower spin-down rate. On the other hand, there is no clear evidence of a variation of the other spectral parameters in the flares after the spin-down rate decreased. Change in $n_H$ may be related to a change in the accretion geometry of the source. However, if there had been an accretion geometry change accompanied with the change in the spin-down rate of the source, we would expect other spectral parameters to vary as well, especially the power law index as in the RXTE observations of SAX J2103.5+4545 (Baykal et al. 2002), and 2S 1417-62 (Inam et al. 2004). New X-ray observations of the source will be useful to monitor any possible changes in the spin-down rate and its consequences.   

{\bf{Acknowledgments}}

We thank anonymous referee for useful comments.

\section*{References}

\noindent{Baykal A., Inam, C., Alpar M.A., in 't Zand J., Strohmayer T., 2001, MNRAS, 327, 1269}

\noindent{Baykal A., Stark M., Swank J., 2002, ApJ, 569, 903}

\noindent{Blondin J.M., Kallman T.R., Fryxell B.A., Taam R.E., 1990, ApJ, 356, 591}

\noindent{Casumano G., Salvo T.D., Burderi L., Orlandini M., Piraino S., Robba N., Santangelo A., 1998, A\& A, 338, L79}

\noindent{Chitnis V.R., Rao A.R., Agrawal P.C., Manchanda R.K., 1993, A\& A, 268, 609}

\noindent{Cook M.C., Page C.G., 1987, MNRAS, 225, 381}

\noindent{Cox N.L., Kaper L., Mokiem M.R., 2005, A\& A, 436, 661}

\noindent{Deeter J.E., Boynton P.E., 1985, in Hayakawa S., Nagase F., eds, Proc. Inuyama Workshop, Timing Studies of X-ray Sources, Nagoya Univ., Nagoya, p.29}

\noindent{Ghosh P., Lamb F.K., 1979, ApJ, 234, 296}

\noindent{Giacconi R., Kellogg E., Gorenstein P., Gursky H., Tananbaum H., 1971, ApJ, 165, L27}

\noindent{Inam S.C., Baykal A., Scott D.M., Finger, M., Swank J., 2004, MNRAS, 349, 173}

\noindent{Jahoda K., Swank J., Giles A.B., Stark M.J., Strohmayer T., Zhang W., 1996, Proc. SPIE, 2808, 59}

\noindent{Leahy D.A., Darbro W., Elsner R.F., Weisskopf M.C., Sutherland P.G., Kahn S., Grindlay J.E., 1983, ApJ, 266, 160}

\noindent{Makishima K., Kawai N., Koyama K., Shibazaki N., Nagase F., Nakagawa M., 1984, PASJ, 36, 679}

\noindent{Makishima K., Ohashi T., Kawai N. et al. 1990, PASJ, 42, 295}

\noindent{Makisihima K., Mihara T., 1992, in Tanaka Y., Koyama K., eds. Proc. XXVIII Yamada Conf., Frontiers of X-ray Astronomy, Universal Academy Press, Tokyo, p. 23}

\noindent{Marshall N., Rickets M.J., 1980, MNRAS, 193, 7}

\noindent{Mihara T., 1995, PhD thesis, Univ. Tokyo}

\noindent{Mukerjee K., Agrawal P.C., Paul B., Rao A.R., Yadav J.S., Seetha S., Kasturirangan K., 2001, ApJ, 548, 368}

\noindent{Roberts M.S.E., Michelson, P.F., Leahy D.A., Hall, T.A., Finley J.P., Cominsky, L.R., Srinivasan, R., 2001, ApJ, 555, 967}

\noindent{in 't Zand J.J.M., Strohmayer T.E., Baykal A., 1997, ApJ, 479, L47}

\noindent{in 't Zand J.J.M., Baykal A., Strohmayer T.E., 1998a, ApJ, 496, 386}

\noindent{in 't Zand J.J.M., Strohmayer T.E., Baykal A., 1998b, Nucl.Phys. B, 69, 224}

\noindent{Valinia A., Marshall F.E., 1998, ApJ, 505, 134}

\noindent{van der Klis M., 1989, in Timing Neutron Stars, ed. H. Ogelman, E.P.J. van den Heuvel (Dordrecht:Kluwer), p.27}

\end{document}